\documentclass[aps,prb, reprint, floatfix, superscriptaddress, showpacs]{revtex4-1}

\usepackage{graphicx} 
\begin{document}

\title{Superconductivity distorted by the coexisting pseudogap in the antinodal region of Bi$_{1.5}$Pb$_{0.55}$Sr$_{1.6}$La$_{0.4}$CuO$_{6+\delta}$: A photon-energy-dependent angle-resolved photoemission study}

\author{M. Hashimoto}
\affiliation{Stanford Synchrotron Radiation Lightsource, SLAC National Accelerator Laboratory, 2575, Sand Hill Road, Menlo Park, California 94025, USA}
\author{R.-H. He}
\affiliation{Stanford Institute for Materials and Energy Sciences, SLAC National Accelerator Laboratory, 2575 Sand Hill Road, Menlo Park, California 94025, USA}
\affiliation{Geballe Laboratory for Advanced Materials, Departments of Physics and Applied Physics, Stanford University, Stanford, California 94305, USA}
\affiliation{Advanced Light Source, Lawrence Berkeley National Lab, Berkeley, California 94720, USA}
\author{I. M. Vishik}
\affiliation{Stanford Institute for Materials and Energy Sciences, SLAC National Accelerator Laboratory, 2575 Sand Hill Road, Menlo Park, California 94025, USA}
\affiliation{Geballe Laboratory for Advanced Materials, Departments of Physics and Applied Physics, Stanford University, Stanford, California 94305, USA}
\author{F. Schmitt}
\affiliation{Stanford Institute for Materials and Energy Sciences, SLAC National Accelerator Laboratory, 2575 Sand Hill Road, Menlo Park, California 94025, USA}
\affiliation{Geballe Laboratory for Advanced Materials, Departments of Physics and Applied Physics, Stanford University, Stanford, California 94305, USA}
\author{R. G. Moore}
\affiliation{Stanford Institute for Materials and Energy Sciences, SLAC National Accelerator Laboratory, 2575 Sand Hill Road, Menlo Park, California 94025, USA}
\affiliation{Geballe Laboratory for Advanced Materials, Departments of Physics and Applied Physics, Stanford University, Stanford, California 94305, USA}
\author{D. H. Lu}
\affiliation{Stanford Synchrotron Radiation Lightsource, SLAC National Accelerator Laboratory, 2575, Sand Hill Road, Menlo Park, California 94025, USA}
\author{Y. Yoshida}
\affiliation{Nanoelectronics Research Institute, AIST, Ibaraki 305-8568, Japan}
\author{H. Eisaki}
\affiliation{Nanoelectronics Research Institute, AIST, Ibaraki 305-8568, Japan}
\author{Z. Hussain}
\affiliation{Advanced Light Source, Lawrence Berkeley National Lab, Berkeley, California 94720, USA}
\author{T. P. Devereaux}
\affiliation{Stanford Institute for Materials and Energy Sciences, SLAC National Accelerator Laboratory, 2575 Sand Hill Road, Menlo Park, California 94025, USA}
\affiliation{Geballe Laboratory for Advanced Materials, Departments of Physics and Applied Physics, Stanford University, Stanford, California 94305, USA}
\author{Z.-X. Shen}
\affiliation{Stanford Institute for Materials and Energy Sciences, SLAC National Accelerator Laboratory, 2575 Sand Hill Road, Menlo Park, California 94025, USA}
\affiliation{Geballe Laboratory for Advanced Materials, Departments of Physics and Applied Physics, Stanford University, Stanford, California 94305, USA}

\date{\today}

\begin{abstract}
The interplay between superconductivity and the pseudogap is an important aspect of cuprate physics. However, the nature of the pseudogap remains controversial, in part because different experiments have suggested different gap functions. Here we present a photon-energy-dependence angle-resolved photoemission spectroscopy (ARPES) study on
Bi$_{1.5}$Pb$_{0.55}$Sr$_{1.6}$La$_{0.4}$CuO$_{6+\delta}$. We find that antinodal ARPES spectra at low photon energies are dominated by background signals which can lead to a misevaluation of the spectral gap size. Once background is properly accounted for, independent of photon energy, the antinodal spectra robustly show two coexisting features at different energies dominantly attributed to the pseudogap and superconductivity, as well as an overall spectral gap which deviates from a simple $d$-wave form. These results support the idea that the spectral gap is distorted due to the competition between the pseudogap and superconductivity.
\end{abstract}


\maketitle
\section{Introduction}
In the high-temperature (T$_c$) cuprate superconductors, the nature of the pseudogap, the mysterious state found above T$_c$, is one of the greatest unresolved problems \cite{Timusk99}.  Owing to their close proximity on the phase diagram, the pseudogap is believed to have a close relation to superconductivity, but whether it represents precursor Cooper pairs or a distinct order competing with superconductivity is a continuing topic of discussion. Angle-resolved photoemission spectroscopy (ARPES) is an important technique for addressing this question \cite{Damascelli03}, because ARPES can study the cuprates in a momentum-resolved manner, for instance, by specifically studying along the Brillouin zone boundary (the antinodal region) where the pseudogap is most pronounced.

Important progress has been made recently by observing the signatures of a second phase transition at T* in addition to superconducting transition at T$_c$ in Bi$_{1.5}$Pb$_{0.55}$Sr$_{1.6}$La$_{0.4}$CuO$_{6+\delta}$ (Pb-Bi2201)\cite{He11}. Additionally, two distinct spectral features are observed in the antinodal region.  The higher energy feature is associated mainly with the pseudogap having broken-symmetry nature \cite{Hashimoto10}, while the lower energy feature is associated mainly with superconductivity. These experiments suggest that the pseudogap cannot be interpreted in terms of precursor Cooper pairs, and instead, point to coexisting order in the high-T$_c$ ground state. The interplay between the two energy scales naturally explains the gap function typically observed with moderate photon energies ($h\nu \sim$20 eV) where the antinodal gap ($\Delta _{AN}\sim$30 meV) strongly deviates from the near-nodal canonical $d$-wave form, $\Delta _0| \cos k_x- \cos k_y|/2 $ with $\Delta _0\sim$15 meV \cite{Kondo07, Hashimoto09, Kondo09, He11, Hashimoto11, Okada11}. However, the form of the gap function is still under debate as other ARPES experiments, particularly those performed at low photon energies ($h\nu <$10 eV), have reported a simple $d$-wave gap function all around the Fermi surface with $\Delta _0\sim$15 meV, and argued that it has a solely superconducting origin \cite{Ma08, Meng09}.  Also, a recent ARPES study on Bi2201 using two different photon energies (21.218 eV and 8.437 eV) has reported the coexistence of two gaps in the antinodal region at $\sim$15 meV (8.437 eV) and $\sim$30 meV (21.218 eV), which have been assigned to superconductivity and the ``soft'' pseudogap which does not distort the simple $d$-wave form of the spectral gap, respectively \cite{Nakayama11}. The momentum dependence of the low temperature gap is debated in other cuprate systems as well \cite{Valla06, Kanigel06, Shi08, Shi09, Tanaka06, Lee07, Terashima07, He09, Yoshida09}, though disagreement is most pronounced about Bi2201. It is important to investigate the photon energy dependence systematically to further understand the nature of the pseudogap.

In this paper, we chose Pb-Bi2201 to study this problem at various incident photon energies from 7 to 22.7 eV. We find that contradictory reports arise because ARPES spectra obtained at lower photon energies are generally contaminated with strong photoemission background signals. Nevertheless, after background subtraction, the spectra show a robust coexistence of two energy scales independent of photon energy. Also, the lower energy gap size smoothly deviates from the simple $d$-wave extension of the near nodal gap. This is consistent with reports using moderate photon energy $h\nu \sim$ 20 eV, but inconsistent with reports using very low incident photon energy. This result strongly supports previous assignments of the two energy scales to the competing pseudogap and superconducting orders, and highlights a strong distortion of the antinodal superconductivity by the coexisting pseudogap order \cite{He11}. 

\section{Experiment}
A high-quality single crystal of optimally doped Pb-Bi2201 was grown by the floating zone method, and the hole concentration was optimized by sample annealing in $N_2$ flow \cite{He11}. T$_c$ was $\sim$  38 K and the pseudogap temperature T* was  $\sim$ 142 K \cite{He11}. Pb doping  significantly suppresses the super-lattice modulation which creates replica of the dispersion in ARPES spectra. ARPES measurements were done at BL 5-4 in SSRL with a Scienta R4000 analyzer. The photon energy $h\nu $ was varied from 8 eV to 22.7 eV. The energy resolution was 7 -- 15 meV. For all photon energies, the polarization was linear and perpendicular to the measured cut. ARPES measurements using a 7 eV laser and circular polarization were also performed with a Scienta 2002 analyzer. The energy resolution was $\sim$3 meV. The Fermi level was calibrated using a gold sample at each photon energy. The measurement temperature was $\sim$7 K and the vacuum pressure was better than 4*10$^{-11}$ Torr.

\begin{figure}
\includegraphics[width=\linewidth]{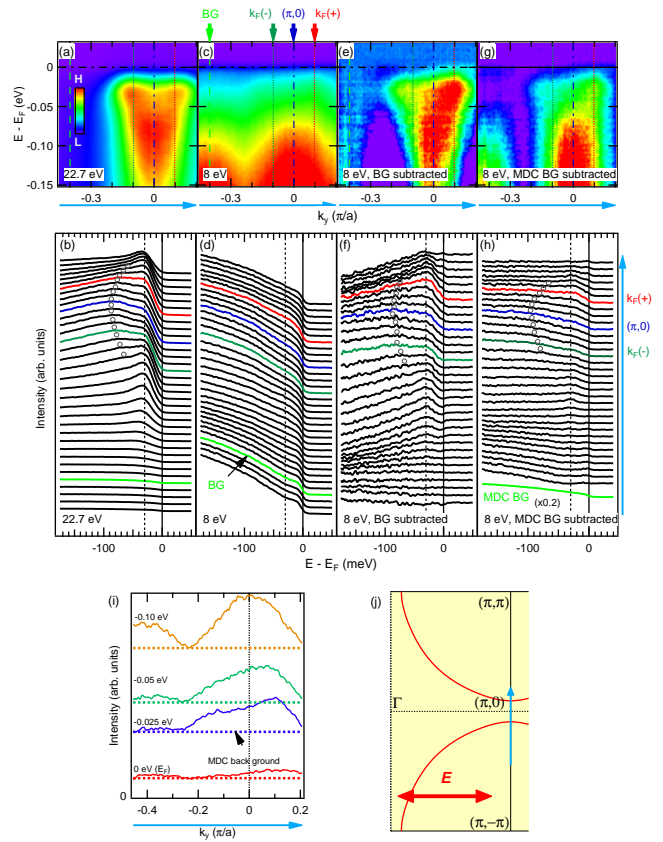}
\caption{Recovery of the intrinsic spectral features by background subtraction analysis in the superconducting state (7 K) of Pb-Bi2201 (a),(b) Image plot and EDCs for the raw ARPES spectra along the antinodal cut taken at $h\nu $ = 22.7 eV. (c),(d) Image plot and EDCs for the raw ARPES spectra at $h\nu $ = 8 eV. (e),(f) Image plot and EDCs after the background subtraction for the 8 eV spectra. (g),(h) Image plot and EDCs after the MDC background subtraction for the 8 eV spectra. The momenta for the background, k$_F$, ($\pi$,0) are indicated by the lines, arrows and colored EDCs. Vertical dashed lines in (b),(d),(f) and (h) at 30 meV and open circles for the dispersion at higher binding energy are guide-to-the-eyes. (i) MDCs at selected energies for the raw spectra ($h\nu $ = 8 eV). Solid curves are the raw spectra and dashed lines are the subtracted background, which are used to obtain the background spectrum shown in (h).(j) Schematic Fermi surface with the measured cut and incident photon polarization. }
\label{Fig1}
\end{figure}

\section{Results and Discussion}
We observe strong contrast between the raw ARPES spectra at the moderate ($\sim$20 eV) and low photon energies ($<$10 eV) obtained within the same experiment on the same sample. Figures \ref{Fig1}(a)-(d) show the antinodal ARPES spectra along ($\pi$,-$\pi$)-($\pi$,0)-($\pi$,$\pi$) taken at two different photon energies, 22.7 eV and 8 eV. At 22.7 eV, [Fig. \ref{Fig1}(a)], the spectral intensity appears as an inverted triangle centered at ($\pi$,0) with a gap feature. The background intensity well away from the Fermi momenta (k$_F$), estimated from the momentum distribution curve (MDC) at E$_F$ \cite{Hashimoto11, He11}, is negligible in the image. The intensity maxima in the energy distribution curves (EDCs) in Fig. \ref{Fig1} (b) show a dispersion down to $\sim$ 85 meV  at ($\pi$,0), which coexists with a shoulder feature at $\sim$ 30 meV that shows little dispersion. These coexisting energy features at 22.7 eV are consistent with previous reports at the same photon energy, and suggest the coexistence of superconductivity and a competing pseudogap \cite{He11}. Also, the energy position of the shoulder feature at $\sim$ 30 meV strongly deviates from the canonical $d$-wave gap function extrapolated from near nodal data which gives $\Delta _0\sim$ 15 meV \cite{Kondo07, Hashimoto09, Kondo09, He11, Hashimoto11, Okada11, Ma08, Meng09, Nakayama11}. 

In strong contrast, the ARPES spectra taken at a low photon energy (8 eV) are almost featureless, as shown in Figs. \ref{Fig1}(c) and (d). The intensity near E$_F$ is smaller and monotonically increases towards the higher binding energy. The EDCs show almost no momentum dependence along the measured cut. One can no longer clearly define a dispersing feature at higher binding energy and the shoulder feature at $\sim$ 30 meV [dashed line in panel (d)] as in the 22.7 eV data. Instead, all the EDCs show a small and broad edge at much lower energy, 10 -- 15 meV, consistent with the previous ARPES studies using lower photon energies \cite{Ma08, Meng09, Nakayama11}. 

\begin{figure*}
\includegraphics[width=\linewidth]{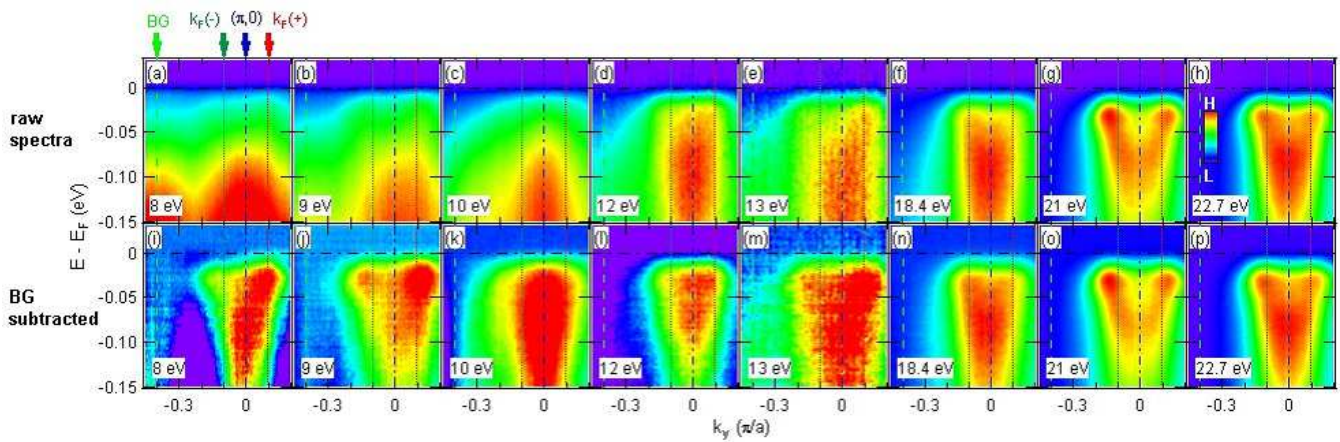}
\caption{Photon energy dependence of the antinodal ARPES spectra before and after the background subtraction analysis. (a)-(l) Image plots with $h\nu $ = 8 -- 22.7 eV. The momenta for the background, k$_F$, ($\pi$,0) are indicated by the lines and arrows.}
\label{Fig2}
\end{figure*}

There are two factors that can make the ARPES spectra look different from the single-particle spectral function. One is matrix elements and the other is background. The relative influence of each depends on incident photon energy and polarization \cite{Damascelli03}. On face value, it is difficult to explain the difference between 22.7 eV and 8 eV spectra only by matrix elements because the spectra are completely different. In the low photon energy spectra [Fig. \ref{Fig1}(d)] in the superconducting state, the edge feature at 10 -- 15 meV expands to an extremely wide momentum range on the Brillouin zone boundary. This anomalously flat feature in momentum space is difficult to interpret as the intrinsic dispersion in the superconducting state. This is because in the true normal state above T* in the same Pb-Bi2201 system \cite{Hashimoto10, He11}, a simple parabolic dispersion has been observed, and one can easily calculate how the band would look after the opening of a 15 meV superconducting gap \cite{Hashimoto10}. Also, the band bottom at($\pi$, 0) in the true normal state is at $\sim$ 20 meV \cite{Hashimoto10, He11}, and the opening of a superconducting gap should shift the band bottom further away from E$_F$. Thus, one expects the lowest energy feature at ($\pi$, 0) to be $>$20 meV. However, the lowest energy feature is at 10 -- 15 meV, smaller than the 20 meV band bottom in the true normal state. Further, the coexistence of the 10 -- 15 meV feature and the 30 meV feature needs to be understood if the 10 -- 15 meV feature is intrinsic. However, it has been indicated that there is no theory supporting such coexistence (Ref. \onlinecite{Nakayama11}). These unusual behaviors make it difficult to interpret the 10 -- 15 meV feature as the intrinsic gap feature. 

Instead, the anomalous observation at 8 eV can be better explained by considering background effects, which can become a substantial part of the spectrum if the intrinsic signal intensity is strongly suppressed by matrix element effects. We applied simple, conventional momentum-independent background subtraction analysis, which has previously been applied to ARPES. Note that the momentum independent background has been assumed in MDC fitting of the ARPES spectra in general. As shown in Figs. 1(e) and (f), we assumed the EDC at k$_y$ = 0.4$\pm $0.01 ($\pi $/a), far from relevant features, as the momentum independent background EDC, and subtracted the background EDC from all the raw EDCs, similar to previous ARPES studies \cite{Kaminski04, Rosch05, Kim07}. After the background EDC subtraction, the ARPES images at 8 and 22.7 eV are more similar. The EDCs [Figs. \ref{Fig1}(b) and (f)] also exhibit quantitative similarity. A gap feature at $\sim$ 30 meV is clear around ($\pi $, 0), and becomes weaker away from k$_F$. The higher energy feature located at $\sim$ 85 meV at ($\pi $,0) seems to be dispersive. In addition, the feature which exists at 10 -- 15 meV in the raw spectra disappears after background subtraction. These results are robust against the momentum chosen for the background EDC if the momentum is sufficiently away from k$_F$ ($\mid $k$_y\mid >$0.25). Some negative values in the background-subtracted EDCs at higher binding energy suggest that the background is overestimated at higher binding energy, which could be due to the effect of shadow bands \cite{Nakayama06, King11}. Polarization dependence in a future study might be important for further understanding of such effects and to obtain more accurate background intensity at higher binding energy. However, this does not affect the features closer to E$_F$, which are of interest in the present study.

To further confirm the robustness of the result, we applied a different background subtraction to the same data set as shown in Figs. 1(g)-(i). Here we assumed the smallest intensity for the momentum distribution curve (MDC) at each energy as background signal [Fig. 1(i)]. The background spectrum obtained from the MDC analysis is shown at the bottom of Fig. 1(h), which is qualitatively similar to the background spectrum in Fig. 1(d). The image and EDCs after the MDC background subtraction are shown in Figs. 1(g) and (h), respectively. One can see that all the energy features of interest are consistent with the result of the background subtraction in Figs. 1(e) and (f). In particular, the lowest energy feature after the MDC background subtraction is at $\sim$ 30 meV and the 10 -- 15 meV edge feature is absent. This MDC background subtraction also avoids the unphysical negative value in the background-subtracted EDCs from the method used in Figs. 1(e) and (f). Because we do not find difference in our conclusion between the two different background subtractions, we present the result of the simplest background subtraction [Fig. 1(e) and (f)] in the rest of this study.

We emphasize that similar background subtraction procedures have been applied to ARPES spectra successfully \cite{Kaminski04, Rosch05, Kim07}, with the background mainly attributed to electrons which have lost their momentum information in the solid. The background signal has been especially well studied in Ref.~\onlinecite{Kaminski04} and it has been shown that the background spectra themselves are gapped in a gapped state, reflecting ``average'' or ``integrated'' gap size (with weighting from different original momenta given by matrix elements), and have clear temperature dependence. This gives further evidence that the momentum independent 10 -- 15 meV edge feature in the raw spectra, which is clearer away from k$_F$ and disappears after the analysis, is a background signal reflecting the averaged gap size in momentum space. The spectral lineshapes around 10 meV near k$_F$ look slightly different from those away from k$_F$ because the weak and broad 30 meV intrinsic feature exists on top of the 10 -- 15 meV background feature. As demonstrated here, a simple background subtraction is sufficient to detect intrinsic features. This also suggests that matrix elements suppress the intrinsic feature strongly in the antinodal region at low photon energies, different from the nodal region \cite{Vishik10}.  

Figure \ref{Fig2} shows the detailed photon energy dependence of the antinodal cut of Pb-Bi2201 from 8 eV to 22.7 eV, before and after the background subtraction. In the raw data, there is a general trend that the spectra become more broad and featureless at lower photon energy. One might naively assume that different photon energies detect different electronic states from these raw images. However, after the background subtraction, the spectra look more similar. The inverted-triangle-like shape around ($\pi $,0) can be robustly observed at different photon energies. This further validates that the the analysis can eliminate the background signals and illuminate the intrinsic spectral function. 

\begin{figure}
\includegraphics[width=\linewidth]{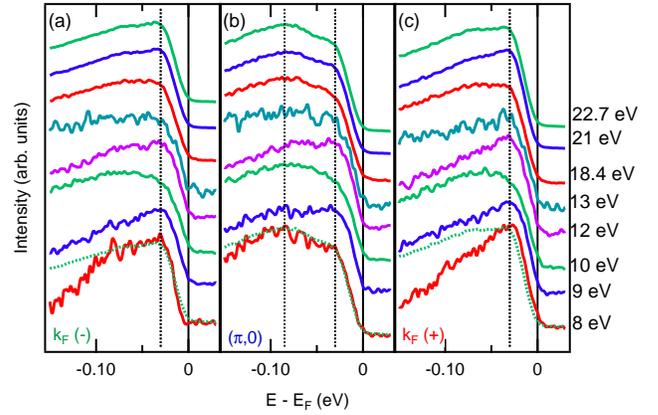}
\caption{Photon energy dependence of the EDCs at k$_F$ and ($\pi$,0) after the background subtraction analysis. (a) EDCs at k$_F$(-) (b) EDCs at ($\pi$,0)(c) EDCs at k$_F$(+). Dashed curves overlaid to spectra at $h\nu $ = 8 eV are the 22.7 eV EDCs. Dotted lines at 30 meV and 85 meV are eyeguides for the two spectral features.}
\label{Fig3}
\end{figure}

The background subtracted EDCs at characteristic momenta, k$_F$ and ($\pi$,0), are displayed in Fig. \ref{Fig3}. All the EDCs robustly show the coexistence of two energy features at $\sim$ 30 meV and $\sim$ 85 meV at ($\pi$,0) \cite{Hashimoto10, He11}. The EDCs at 22.7 eV and 8 eV exhibit similar energy scales as overlaid in the figure. This suggests that the coexistence of two order parameters, the pseudogap and superconductivity \cite{He11}, is robust against photon energy. This coexistence is different from the coexistence of $\sim$15 and $\sim$30 meV gaps in the antinode discussed in Ref.~\onlinecite{Nakayama11} where the two gaps coexist without interaction. We note that, in the intermediate photon energies, 14--17 eV, strong shadow band effects \cite{Nakayama06, King11} make the evaluation of the background difficult (not shown). However, importantly, the 30 meV gap feature remains robust even without any background subtraction at these photon energies.

\begin{figure}
\includegraphics[width=\linewidth]{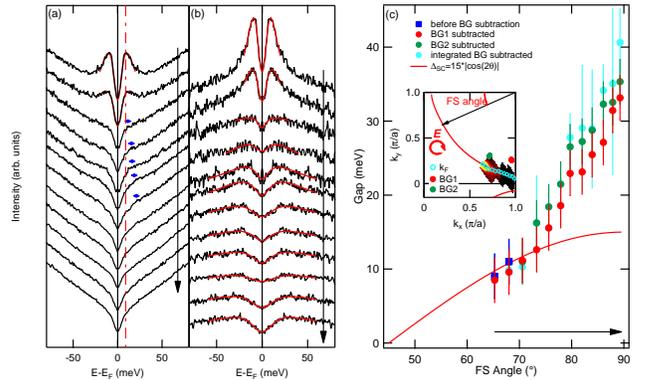}
\caption{Deviation from the canonical $d$-wave form in laser ARPES data ($h\nu$ = 7 eV) (a) Symmetrized raw EDCs on the Fermi surface. Blue markers are the guides to eyes for the weak intrinsic gap feature at intermediate momenta, and the dashed line at 10 meV is the guide to eyes for the comparison between the spectrum at the antinode and at different momenta.(b) Symmetrized EDCs after the background subtraction. Fitted curves are overlaid. (c) Gap function obtained from symmetrized EDCs before and after background subtraction. Integrated background EDC is obtained by averaging the entire measured spectra. Inset shows the Fermi surface intensity plot. Fermi momenta and background EDC momenta (BG1 and BG2) are also shown.}
\label{Fig5}
\end{figure}

We study the gap function using 7 eV laser ARPES to further confirm the strong background signal in the antinode and the deviation of the gap function from a simple $d$-wave. The symmetrized EDCs along the FS are plotted in Fig. \ref{Fig5}(a). The measurement covers from $\sim$20 $^{\circ}$ away from the node in Fermi surface angle, where superconducting quasiparticle peaks are clearly visible, to the antinode, where they are not, as mapped in Fig. \ref{Fig5}(d). Due to the experimental setup, the entire FS is not covered. Without any background subtraction, the peak feature quickly becomes weaker towards the antinode. Near the node, one can fit the symmetrized EDCs to a commonly used phenomenological model \cite{Norman98}, as shown in Fig. \ref{Fig5}(a), but moving towards the antinode, it soon becomes impossible to fit without assuming a background signal. However, in the intermediate momenta, one can still see a weak intrinsic gap feature at higher energy than the 10 -- 15 meV feature at the antinode in the raw spectrum. If the 10 -- 15 meV feature at the antinode is intrinsic, the gap function becomes non-monotonic and this is unlikely.

We applied the same background subtraction method to these spectra and the results are shown in Fig. \ref{Fig5}(b). Here we assumed a common background EDC for the entire FS, as indicated as BG1 in Fig. \ref{Fig5}(d). Note that we measured a cut for the background EDC near the antinode due to the limitation of the acceptance angle of the analyzer. After the background subtraction, as shown in Fig. \ref{Fig5}(b), all the symmetrized EDCs show a peak feature which allows us to fit the EDCs to the phenomenological model, and the fitting results in red are overlaid to the fitted spectra. The gap magnitude derived in this manner is plotted as a function of momentum in Fig. \ref{Fig5}(c). In the intermediate momentum region ($\sim$20 $^{\circ}$ away from the node in FS angle), the gap magnitudes are consistent with a $d$-wave gap function with $\Delta _0\sim$15 meV, both in the raw and background-subtracted EDCs, suggesting that the background signal is not very strong. However, approaching the antinode, the gap function derived from the background-subtracted spectra clearly and smoothly deviates from a simple $d$-wave form, and shows maximum gap size of $\sim$30 meV at the antinode. The EDCs are nearly momentum independent in the entire Brillouin zone when sufficiently away from the dispersion. We found that the background EDC (BG2) obtained at a different momentum [Fig. \ref{Fig5}(d)] results in a consistent gap function as shown in Fig. \ref{Fig5}(c). Further, a similar gap function can be obtained when we choose an EDC obtained by averaging the entire measured momenta as background [Fig. \ref{Fig5}(c)], supporting the idea that the background is the result of scattered electrons that lost their momentum information. Also, note that the conclusion does not seem to be strongly affected by the incident photon polarization (linear polarization for Fig. 1 -- 3 and circular polarization for Fig. 4), but future polarization-dependence study might be important to further understand the nature of the background. This analysis further confirms the strong background signal in the antinodal region and the photon-energy-independent gap function which shows a deviation from a simple $d$-wave in the antinodal region. 

There might remain a small possibility that the contradictory low photon energy reports suggesting a simple $d$-wave gap \cite{Ma08, Meng09, Nakayama11} are due to subtle material, doping or experimental condition differences. However, the present result suggests that these contradictory results could be reconciled by considering a strong background signal. The smooth momentum dependence of the gap function reported with low photon energies could reflect the intrinsic gap feature being overwhelmed gradually by the background signal with a gap-like structure at 10 -- 15 meV, going from the node to the antinode. Taking into account that the other single-layer cuprates also show a deviation from the canonical $d$-wave form in the antinodal region \cite{He09, Yoshida09}, it is advisable to revisit the low photon energy ARPES studies by taking into account the strong contribution of the background signals which could contaminate the intrinsic gap signal.

In scanning tunneling spectroscopy (STS), the coexistence of two energy scales has been suggested \cite{Ma08, McElroy05, Pushp09}, similar to ARPES. However, how they coexist in momentum space is not clear from STS because the STS spectra are momentum integrated with matrix elements \cite{Hashimoto11}. Nevertheless, an attempt to extract the momentum dependence of the gap has suggested that the gap function deviates from a simple $d$-wave form in the antinodal region \cite{Pushp09}, supporting the present result. 

\section{Conclusion}
In summary, our photon energy dependence study revealed that the antinodal ARPES spectra are strongly influenced by background intensity at low photon energy. With the background subtraction analysis, we showed the robustness the two coexisting energy scales for pseudogap and superconductivity at the antinode. Thus, the gap function deviates from a simple $d$-wave form at the antinode, and the data are inconsistent with a simple extension of the $d$-wave nodal gap to the antinode. Instead, our results suggest that the spectral gap is distorted due to the competition between the pseudogap, a distinct order from superconductivity with a broken-symmetry nature, and superconductivity.  

\section{Acknowledgment}
This work is supported by the Department of Energy, Office of Basic Energy Science, Division of  Materials Science.

%\bibliography{hndep}

\begin{thebibliography}{30}%
\makeatletter
\providecommand \@ifxundefined [1]{%
 \@ifx{#1\undefined}
}%
\providecommand \@ifnum [1]{%
 \ifnum #1\expandafter \@firstoftwo
 \else \expandafter \@secondoftwo
 \fi
}%
\providecommand \@ifx [1]{%
 \ifx #1\expandafter \@firstoftwo
 \else \expandafter \@secondoftwo
 \fi
}%
\providecommand \natexlab [1]{#1}%
\providecommand \enquote  [1]{``#1''}%
\providecommand \bibnamefont  [1]{#1}%
\providecommand \bibfnamefont [1]{#1}%
\providecommand \citenamefont [1]{#1}%
\providecommand \href@noop [0]{\@secondoftwo}%
\providecommand \href [0]{\begingroup \@sanitize@url \@href}%
\providecommand \@href[1]{\@@startlink{#1}\@@href}%
\providecommand \@@href[1]{\endgroup#1\@@endlink}%
\providecommand \@sanitize@url [0]{\catcode `\\12\catcode `\$12\catcode
  `\&12\catcode `\#12\catcode `\^12\catcode `\_12\catcode `\%12\relax}%
\providecommand \@@startlink[1]{}%
\providecommand \@@endlink[0]{}%
\providecommand \url  [0]{\begingroup\@sanitize@url \@url }%
\providecommand \@url [1]{\endgroup\@href {#1}{\urlprefix }}%
\providecommand \urlprefix  [0]{URL }%
\providecommand \Eprint [0]{\href }%
\providecommand \doibase [0]{http://dx.doi.org/}%
\providecommand \selectlanguage [0]{\@gobble}%
\providecommand \bibinfo  [0]{\@secondoftwo}%
\providecommand \bibfield  [0]{\@secondoftwo}%
\providecommand \translation [1]{[#1]}%
\providecommand \BibitemOpen [0]{}%
\providecommand \bibitemStop [0]{}%
\providecommand \bibitemNoStop [0]{.\EOS\space}%
\providecommand \EOS [0]{\spacefactor3000\relax}%
\providecommand \BibitemShut  [1]{\csname bibitem#1\endcsname}%
\let\auto@bib@innerbib\@empty
%</preamble>
\bibitem [{\citenamefont {Timusk}\ and\ \citenamefont
  {Statt}(1999)}]{Timusk99}%
  \BibitemOpen
  \bibfield  {author} {\bibinfo {author} {\bibfnamefont {T.}~\bibnamefont
  {Timusk}}\ and\ \bibinfo {author} {\bibfnamefont {B.}~\bibnamefont {Statt}},\
  }\href@noop {} {\bibfield  {journal} {\bibinfo  {journal} {Rep. Prog. Phys.}\
  }\textbf {\bibinfo {volume} {62}},\ \bibinfo {pages} {61} (\bibinfo {year}
  {1999})}\BibitemShut {NoStop}%
\bibitem [{\citenamefont {Damascelli}\ \emph {et~al.}(2003)\citenamefont
  {Damascelli}, \citenamefont {Hussain},\ and\ \citenamefont
  {Shen}}]{Damascelli03}%
  \BibitemOpen
  \bibfield  {author} {\bibinfo {author} {\bibfnamefont {A.}~\bibnamefont
  {Damascelli}}, \bibinfo {author} {\bibfnamefont {Z.}~\bibnamefont {Hussain}},
  \ and\ \bibinfo {author} {\bibfnamefont {Z.~X.}\ \bibnamefont {Shen}},\
  }\href@noop {} {\bibfield  {journal} {\bibinfo  {journal} {Rev. Mod. Phys.}\
  }\textbf {\bibinfo {volume} {75}},\ \bibinfo {pages} {473} (\bibinfo {year}
  {2003})}\BibitemShut {NoStop}%
\bibitem [{\citenamefont {He}\ \emph {et~al.}(2011)\citenamefont {He},
  \citenamefont {Hashimoto}, \citenamefont {Karapetyan}, \citenamefont
  {Koralek}, \citenamefont {Hinton}, \citenamefont {Testaud}, \citenamefont
  {Nathan}, \citenamefont {Yoshida}, \citenamefont {Yao}, \citenamefont
  {Tanaka}, \citenamefont {Meevasana}, \citenamefont {Moore}, \citenamefont
  {Lu}, \citenamefont {Mo}, \citenamefont {Ishikado}, \citenamefont {Eisaki},
  \citenamefont {Hussain}, \citenamefont {Devereaux}, \citenamefont {Kivelson},
  \citenamefont {Orenstein}, \citenamefont {Kapitulnik},\ and\ \citenamefont
  {Shen}}]{He11}%
  \BibitemOpen
  \bibfield  {author} {\bibinfo {author} {\bibfnamefont {R.~H.}\ \bibnamefont
  {He}}, \bibinfo {author} {\bibfnamefont {M.}~\bibnamefont {Hashimoto}},
  \bibinfo {author} {\bibfnamefont {H.}~\bibnamefont {Karapetyan}}, \bibinfo
  {author} {\bibfnamefont {J.~D.}\ \bibnamefont {Koralek}}, \bibinfo {author}
  {\bibfnamefont {J.~P.}\ \bibnamefont {Hinton}}, \bibinfo {author}
  {\bibfnamefont {J.~P.}\ \bibnamefont {Testaud}}, \bibinfo {author}
  {\bibfnamefont {V.}~\bibnamefont {Nathan}}, \bibinfo {author} {\bibfnamefont
  {Y.}~\bibnamefont {Yoshida}}, \bibinfo {author} {\bibfnamefont
  {H.}~\bibnamefont {Yao}}, \bibinfo {author} {\bibfnamefont {K.}~\bibnamefont
  {Tanaka}}, \bibinfo {author} {\bibfnamefont {W.}~\bibnamefont {Meevasana}},
  \bibinfo {author} {\bibfnamefont {R.~G.}\ \bibnamefont {Moore}}, \bibinfo
  {author} {\bibfnamefont {D.~H.}\ \bibnamefont {Lu}}, \bibinfo {author}
  {\bibfnamefont {S.~K.}\ \bibnamefont {Mo}}, \bibinfo {author} {\bibfnamefont
  {M.}~\bibnamefont {Ishikado}}, \bibinfo {author} {\bibfnamefont
  {H.}~\bibnamefont {Eisaki}}, \bibinfo {author} {\bibfnamefont
  {Z.}~\bibnamefont {Hussain}}, \bibinfo {author} {\bibfnamefont {T.~P.}\
  \bibnamefont {Devereaux}}, \bibinfo {author} {\bibfnamefont {S.~A.}\
  \bibnamefont {Kivelson}}, \bibinfo {author} {\bibfnamefont {J.}~\bibnamefont
  {Orenstein}}, \bibinfo {author} {\bibfnamefont {A.}~\bibnamefont
  {Kapitulnik}}, \ and\ \bibinfo {author} {\bibfnamefont {Z.~X.}\ \bibnamefont
  {Shen}},\ }\href@noop {} {\bibfield  {journal} {\bibinfo  {journal}
  {Science}\ }\textbf {\bibinfo {volume} {331}},\ \bibinfo {pages} {1579}
  (\bibinfo {year} {2011})}\BibitemShut {NoStop}%
\bibitem [{\citenamefont {Hashimoto}\ \emph {et~al.}(2010)\citenamefont
  {Hashimoto}, \citenamefont {He}, \citenamefont {Tanaka}, \citenamefont
  {Testaud}, \citenamefont {Meevasana}, \citenamefont {Moore}, \citenamefont
  {Lu}, \citenamefont {Yao}, \citenamefont {Yoshida}, \citenamefont {Eisaki},
  \citenamefont {Devereaux}, \citenamefont {Hussain},\ and\ \citenamefont
  {Shen}}]{Hashimoto10}%
  \BibitemOpen
  \bibfield  {author} {\bibinfo {author} {\bibfnamefont {M.}~\bibnamefont
  {Hashimoto}}, \bibinfo {author} {\bibfnamefont {R.~H.}\ \bibnamefont {He}},
  \bibinfo {author} {\bibfnamefont {K.}~\bibnamefont {Tanaka}}, \bibinfo
  {author} {\bibfnamefont {J.~P.}\ \bibnamefont {Testaud}}, \bibinfo {author}
  {\bibfnamefont {W.}~\bibnamefont {Meevasana}}, \bibinfo {author}
  {\bibfnamefont {R.~G.}\ \bibnamefont {Moore}}, \bibinfo {author}
  {\bibfnamefont {D.~H.}\ \bibnamefont {Lu}}, \bibinfo {author} {\bibfnamefont
  {H.}~\bibnamefont {Yao}}, \bibinfo {author} {\bibfnamefont {Y.}~\bibnamefont
  {Yoshida}}, \bibinfo {author} {\bibfnamefont {H.}~\bibnamefont {Eisaki}},
  \bibinfo {author} {\bibfnamefont {T.~P.}\ \bibnamefont {Devereaux}}, \bibinfo
  {author} {\bibfnamefont {Z.}~\bibnamefont {Hussain}}, \ and\ \bibinfo
  {author} {\bibfnamefont {Z.~X.}\ \bibnamefont {Shen}},\ }\href@noop {}
  {\bibfield  {journal} {\bibinfo  {journal} {Nature Phys.}\ }\textbf {\bibinfo
  {volume} {6}},\ \bibinfo {pages} {414} (\bibinfo {year} {2010})}\BibitemShut
  {NoStop}%
\bibitem [{\citenamefont {Kondo}\ \emph {et~al.}(2007)\citenamefont {Kondo},
  \citenamefont {Takeuchi}, \citenamefont {Kaminski}, \citenamefont {Tsuda},\
  and\ \citenamefont {Shin}}]{Kondo07}%
  \BibitemOpen
  \bibfield  {author} {\bibinfo {author} {\bibfnamefont {T.}~\bibnamefont
  {Kondo}}, \bibinfo {author} {\bibfnamefont {T.}~\bibnamefont {Takeuchi}},
  \bibinfo {author} {\bibfnamefont {A.}~\bibnamefont {Kaminski}}, \bibinfo
  {author} {\bibfnamefont {S.}~\bibnamefont {Tsuda}}, \ and\ \bibinfo {author}
  {\bibfnamefont {S.}~\bibnamefont {Shin}},\ }\href@noop {} {\bibfield
  {journal} {\bibinfo  {journal} {Phys. Rev. Lett.}\ }\textbf {\bibinfo
  {volume} {98}},\ \bibinfo {pages} {267004} (\bibinfo {year}
  {2007})}\BibitemShut {NoStop}%
\bibitem [{\citenamefont {Hashimoto}\ \emph {et~al.}(2009)\citenamefont
  {Hashimoto}, \citenamefont {Yoshida}, \citenamefont {Fujimori}, \citenamefont
  {Lu}, \citenamefont {Shen}, \citenamefont {Kubota}, \citenamefont {Ono},
  \citenamefont {Ishikado}, \citenamefont {Fujita},\ and\ \citenamefont
  {Uchida}}]{Hashimoto09}%
  \BibitemOpen
  \bibfield  {author} {\bibinfo {author} {\bibfnamefont {M.}~\bibnamefont
  {Hashimoto}}, \bibinfo {author} {\bibfnamefont {T.}~\bibnamefont {Yoshida}},
  \bibinfo {author} {\bibfnamefont {A.}~\bibnamefont {Fujimori}}, \bibinfo
  {author} {\bibfnamefont {D.~H.}\ \bibnamefont {Lu}}, \bibinfo {author}
  {\bibfnamefont {Z.~X.}\ \bibnamefont {Shen}}, \bibinfo {author}
  {\bibfnamefont {M.}~\bibnamefont {Kubota}}, \bibinfo {author} {\bibfnamefont
  {K.}~\bibnamefont {Ono}}, \bibinfo {author} {\bibfnamefont {M.}~\bibnamefont
  {Ishikado}}, \bibinfo {author} {\bibfnamefont {K.}~\bibnamefont {Fujita}}, \
  and\ \bibinfo {author} {\bibfnamefont {S.}~\bibnamefont {Uchida}},\
  }\href@noop {} {\bibfield  {journal} {\bibinfo  {journal} {Phys. Rev. B}\
  }\textbf {\bibinfo {volume} {79}},\ \bibinfo {pages} {144517} (\bibinfo
  {year} {2009})}\BibitemShut {NoStop}%
\bibitem [{\citenamefont {Kondo}\ \emph {et~al.}(2009)\citenamefont {Kondo},
  \citenamefont {Khasanov}, \citenamefont {Takeuchi}, \citenamefont
  {Schmalian},\ and\ \citenamefont {Kaminski}}]{Kondo09}%
  \BibitemOpen
  \bibfield  {author} {\bibinfo {author} {\bibfnamefont {T.}~\bibnamefont
  {Kondo}}, \bibinfo {author} {\bibfnamefont {R.}~\bibnamefont {Khasanov}},
  \bibinfo {author} {\bibfnamefont {T.}~\bibnamefont {Takeuchi}}, \bibinfo
  {author} {\bibfnamefont {J.}~\bibnamefont {Schmalian}}, \ and\ \bibinfo
  {author} {\bibfnamefont {A.}~\bibnamefont {Kaminski}},\ }\href@noop {}
  {\bibfield  {journal} {\bibinfo  {journal} {Nature}\ }\textbf {\bibinfo
  {volume} {457}},\ \bibinfo {pages} {296} (\bibinfo {year}
  {2009})}\BibitemShut {NoStop}%
\bibitem [{\citenamefont {Hashimoto}\ \emph {et~al.}(2011)\citenamefont
  {Hashimoto}, \citenamefont {He}, \citenamefont {Testaud}, \citenamefont
  {Meevasana}, \citenamefont {Moore}, \citenamefont {Lu}, \citenamefont
  {Yoshida}, \citenamefont {Eisaki}, \citenamefont {Devereaux}, \citenamefont
  {Hussain},\ and\ \citenamefont {Shen}}]{Hashimoto11}%
  \BibitemOpen
  \bibfield  {author} {\bibinfo {author} {\bibfnamefont {M.}~\bibnamefont
  {Hashimoto}}, \bibinfo {author} {\bibfnamefont {R.~H.}\ \bibnamefont {He}},
  \bibinfo {author} {\bibfnamefont {J.~P.}\ \bibnamefont {Testaud}}, \bibinfo
  {author} {\bibfnamefont {W.}~\bibnamefont {Meevasana}}, \bibinfo {author}
  {\bibfnamefont {R.~G.}\ \bibnamefont {Moore}}, \bibinfo {author}
  {\bibfnamefont {D.~H.}\ \bibnamefont {Lu}}, \bibinfo {author} {\bibfnamefont
  {Y.}~\bibnamefont {Yoshida}}, \bibinfo {author} {\bibfnamefont
  {H.}~\bibnamefont {Eisaki}}, \bibinfo {author} {\bibfnamefont {T.~P.}\
  \bibnamefont {Devereaux}}, \bibinfo {author} {\bibfnamefont {Z.}~\bibnamefont
  {Hussain}}, \ and\ \bibinfo {author} {\bibfnamefont {Z.~X.}\ \bibnamefont
  {Shen}},\ }\href@noop {} {\bibfield  {journal} {\bibinfo  {journal} {Phys.
  Rev. Lett.}\ }\textbf {\bibinfo {volume} {106}},\ \bibinfo {pages} {167003}
  (\bibinfo {year} {2011})}\BibitemShut {NoStop}%
\bibitem [{\citenamefont {Okada}\ \emph {et~al.}(2011)\citenamefont {Okada},
  \citenamefont {Kawaguchi}, \citenamefont {Ohkawa}, \citenamefont {Ishizaka},
  \citenamefont {Takeuchi}, \citenamefont {Shin},\ and\ \citenamefont
  {Ikuta}}]{Okada11}%
  \BibitemOpen
  \bibfield  {author} {\bibinfo {author} {\bibfnamefont {Y.}~\bibnamefont
  {Okada}}, \bibinfo {author} {\bibfnamefont {T.}~\bibnamefont {Kawaguchi}},
  \bibinfo {author} {\bibfnamefont {M.}~\bibnamefont {Ohkawa}}, \bibinfo
  {author} {\bibfnamefont {K.}~\bibnamefont {Ishizaka}}, \bibinfo {author}
  {\bibfnamefont {T.}~\bibnamefont {Takeuchi}}, \bibinfo {author}
  {\bibfnamefont {S.}~\bibnamefont {Shin}}, \ and\ \bibinfo {author}
  {\bibfnamefont {H.}~\bibnamefont {Ikuta}},\ }\href@noop {} {\bibfield
  {journal} {\bibinfo  {journal} {Phys. Rev. B}\ }\textbf {\bibinfo {volume}
  {83}},\ \bibinfo {pages} {104502} (\bibinfo {year} {2011})}\BibitemShut
  {NoStop}%
\bibitem [{\citenamefont {Ma}\ \emph {et~al.}(2008)\citenamefont {Ma},
  \citenamefont {Pan}, \citenamefont {Niestemski}, \citenamefont {Neupane},
  \citenamefont {Xu}, \citenamefont {Richard}, \citenamefont {Nakayama},
  \citenamefont {Sato}, \citenamefont {Takahashi}, \citenamefont {Luo},
  \citenamefont {Fang}, \citenamefont {Wen}, \citenamefont {Wang},
  \citenamefont {Ding},\ and\ \citenamefont {Madhavan}}]{Ma08}%
  \BibitemOpen
  \bibfield  {author} {\bibinfo {author} {\bibfnamefont {J.~H.}\ \bibnamefont
  {Ma}}, \bibinfo {author} {\bibfnamefont {Z.~H.}\ \bibnamefont {Pan}},
  \bibinfo {author} {\bibfnamefont {F.~C.}\ \bibnamefont {Niestemski}},
  \bibinfo {author} {\bibfnamefont {M.}~\bibnamefont {Neupane}}, \bibinfo
  {author} {\bibfnamefont {Y.~M.}\ \bibnamefont {Xu}}, \bibinfo {author}
  {\bibfnamefont {P.}~\bibnamefont {Richard}}, \bibinfo {author} {\bibfnamefont
  {K.}~\bibnamefont {Nakayama}}, \bibinfo {author} {\bibfnamefont
  {T.}~\bibnamefont {Sato}}, \bibinfo {author} {\bibfnamefont {T.}~\bibnamefont
  {Takahashi}}, \bibinfo {author} {\bibfnamefont {H.~Q.}\ \bibnamefont {Luo}},
  \bibinfo {author} {\bibfnamefont {L.}~\bibnamefont {Fang}}, \bibinfo {author}
  {\bibfnamefont {H.~H.}\ \bibnamefont {Wen}}, \bibinfo {author} {\bibfnamefont
  {Z.~Q.}\ \bibnamefont {Wang}}, \bibinfo {author} {\bibfnamefont
  {H.}~\bibnamefont {Ding}}, \ and\ \bibinfo {author} {\bibfnamefont
  {V.}~\bibnamefont {Madhavan}},\ }\href@noop {} {\bibfield  {journal}
  {\bibinfo  {journal} {Phys. Rev. Lett.}\ }\textbf {\bibinfo {volume} {101}},\
  \bibinfo {pages} {207002} (\bibinfo {year} {2008})}\BibitemShut {NoStop}%
\bibitem [{\citenamefont {Meng}\ \emph {et~al.}(2009)\citenamefont {Meng},
  \citenamefont {Zhang}, \citenamefont {Liu}, \citenamefont {Zhao},
  \citenamefont {Liu}, \citenamefont {Jia}, \citenamefont {Lu}, \citenamefont
  {Dong}, \citenamefont {Wang}, \citenamefont {Zhang}, \citenamefont {Zhou},
  \citenamefont {Zhu}, \citenamefont {Wang}, \citenamefont {Zhao},
  \citenamefont {Xu}, \citenamefont {Chen},\ and\ \citenamefont
  {Zhou}}]{Meng09}%
  \BibitemOpen
  \bibfield  {author} {\bibinfo {author} {\bibfnamefont {J.~Q.}\ \bibnamefont
  {Meng}}, \bibinfo {author} {\bibfnamefont {W.~T.}\ \bibnamefont {Zhang}},
  \bibinfo {author} {\bibfnamefont {G.~D.}\ \bibnamefont {Liu}}, \bibinfo
  {author} {\bibfnamefont {L.}~\bibnamefont {Zhao}}, \bibinfo {author}
  {\bibfnamefont {H.~Y.}\ \bibnamefont {Liu}}, \bibinfo {author} {\bibfnamefont
  {X.~W.}\ \bibnamefont {Jia}}, \bibinfo {author} {\bibfnamefont
  {W.}~\bibnamefont {Lu}}, \bibinfo {author} {\bibfnamefont {X.~L.}\
  \bibnamefont {Dong}}, \bibinfo {author} {\bibfnamefont {G.~L.}\ \bibnamefont
  {Wang}}, \bibinfo {author} {\bibfnamefont {H.~B.}\ \bibnamefont {Zhang}},
  \bibinfo {author} {\bibfnamefont {Y.}~\bibnamefont {Zhou}}, \bibinfo {author}
  {\bibfnamefont {Y.}~\bibnamefont {Zhu}}, \bibinfo {author} {\bibfnamefont
  {X.~Y.}\ \bibnamefont {Wang}}, \bibinfo {author} {\bibfnamefont {Z.~X.}\
  \bibnamefont {Zhao}}, \bibinfo {author} {\bibfnamefont {Z.~Y.}\ \bibnamefont
  {Xu}}, \bibinfo {author} {\bibfnamefont {C.~T.}\ \bibnamefont {Chen}}, \ and\
  \bibinfo {author} {\bibfnamefont {X.~J.}\ \bibnamefont {Zhou}},\ }\href@noop
  {} {\bibfield  {journal} {\bibinfo  {journal} {Phys. Rev. B}\ }\textbf
  {\bibinfo {volume} {79}},\ \bibinfo {pages} {024514} (\bibinfo {year}
  {2009})}\BibitemShut {NoStop}%
\bibitem [{\citenamefont {Nakayama}\ \emph {et~al.}(2011)\citenamefont
  {Nakayama}, \citenamefont {Sato}, \citenamefont {Xu}, \citenamefont {Pan},
  \citenamefont {Richard}, \citenamefont {Ding}, \citenamefont {Wen},
  \citenamefont {Kudo}, \citenamefont {Sasaki}, \citenamefont {Kobayashi},\
  and\ \citenamefont {Takahashi}}]{Nakayama11}%
  \BibitemOpen
  \bibfield  {author} {\bibinfo {author} {\bibfnamefont {K.}~\bibnamefont
  {Nakayama}}, \bibinfo {author} {\bibfnamefont {T.}~\bibnamefont {Sato}},
  \bibinfo {author} {\bibfnamefont {Y.-M.}\ \bibnamefont {Xu}}, \bibinfo
  {author} {\bibfnamefont {Z.-H.}\ \bibnamefont {Pan}}, \bibinfo {author}
  {\bibfnamefont {P.}~\bibnamefont {Richard}}, \bibinfo {author} {\bibfnamefont
  {H.}~\bibnamefont {Ding}}, \bibinfo {author} {\bibfnamefont {H.-H.}\
  \bibnamefont {Wen}}, \bibinfo {author} {\bibfnamefont {K.}~\bibnamefont
  {Kudo}}, \bibinfo {author} {\bibfnamefont {T.}~\bibnamefont {Sasaki}},
  \bibinfo {author} {\bibfnamefont {N.}~\bibnamefont {Kobayashi}}, \ and\
  \bibinfo {author} {\bibfnamefont {T.}~\bibnamefont {Takahashi}},\ }\href@noop
  {} {\bibfield  {journal} {\bibinfo  {journal} {Phys. Rev. B}\ }\textbf
  {\bibinfo {volume} {83}},\ \bibinfo {pages} {224509} (\bibinfo {year}
  {2011})}\BibitemShut {NoStop}%
\bibitem [{\citenamefont {Valla}\ \emph {et~al.}(2006)\citenamefont {Valla},
  \citenamefont {Fedorov}, \citenamefont {Lee}, \citenamefont {Davis},\ and\
  \citenamefont {Gu}}]{Valla06}%
  \BibitemOpen
  \bibfield  {author} {\bibinfo {author} {\bibfnamefont {T.}~\bibnamefont
  {Valla}}, \bibinfo {author} {\bibfnamefont {A.~V.}\ \bibnamefont {Fedorov}},
  \bibinfo {author} {\bibfnamefont {J.}~\bibnamefont {Lee}}, \bibinfo {author}
  {\bibfnamefont {J.~C.}\ \bibnamefont {Davis}}, \ and\ \bibinfo {author}
  {\bibfnamefont {G.~D.}\ \bibnamefont {Gu}},\ }\href@noop {} {\bibfield
  {journal} {\bibinfo  {journal} {Science}\ }\textbf {\bibinfo {volume}
  {314}},\ \bibinfo {pages} {1914} (\bibinfo {year} {2006})}\BibitemShut
  {NoStop}%
\bibitem [{\citenamefont {Kanigel}\ \emph {et~al.}(2006)\citenamefont
  {Kanigel}, \citenamefont {Norman}, \citenamefont {Randeria}, \citenamefont
  {Chatterjee}, \citenamefont {Souma}, \citenamefont {Kaminski}, \citenamefont
  {Fretwell}, \citenamefont {Rosenkranz}, \citenamefont {Shi}, \citenamefont
  {Sato}, \citenamefont {Takahashi}, \citenamefont {Li}, \citenamefont {Raffy},
  \citenamefont {Kadowaki}, \citenamefont {Hinks}, \citenamefont {Ozyuzer},\
  and\ \citenamefont {Campuzano}}]{Kanigel06}%
  \BibitemOpen
  \bibfield  {author} {\bibinfo {author} {\bibfnamefont {A.}~\bibnamefont
  {Kanigel}}, \bibinfo {author} {\bibfnamefont {M.~R.}\ \bibnamefont {Norman}},
  \bibinfo {author} {\bibfnamefont {M.}~\bibnamefont {Randeria}}, \bibinfo
  {author} {\bibfnamefont {U.}~\bibnamefont {Chatterjee}}, \bibinfo {author}
  {\bibfnamefont {S.}~\bibnamefont {Souma}}, \bibinfo {author} {\bibfnamefont
  {A.}~\bibnamefont {Kaminski}}, \bibinfo {author} {\bibfnamefont {H.~M.}\
  \bibnamefont {Fretwell}}, \bibinfo {author} {\bibfnamefont {S.}~\bibnamefont
  {Rosenkranz}}, \bibinfo {author} {\bibfnamefont {M.}~\bibnamefont {Shi}},
  \bibinfo {author} {\bibfnamefont {T.}~\bibnamefont {Sato}}, \bibinfo {author}
  {\bibfnamefont {T.}~\bibnamefont {Takahashi}}, \bibinfo {author}
  {\bibfnamefont {Z.~Z.}\ \bibnamefont {Li}}, \bibinfo {author} {\bibfnamefont
  {H.}~\bibnamefont {Raffy}}, \bibinfo {author} {\bibfnamefont
  {K.}~\bibnamefont {Kadowaki}}, \bibinfo {author} {\bibfnamefont
  {D.}~\bibnamefont {Hinks}}, \bibinfo {author} {\bibfnamefont
  {L.}~\bibnamefont {Ozyuzer}}, \ and\ \bibinfo {author} {\bibfnamefont
  {J.~C.}\ \bibnamefont {Campuzano}},\ }\href@noop {} {\bibfield  {journal}
  {\bibinfo  {journal} {Nature Phys.}\ }\textbf {\bibinfo {volume} {2}},\
  \bibinfo {pages} {447} (\bibinfo {year} {2006})}\BibitemShut {NoStop}%
\bibitem [{\citenamefont {Shi}\ \emph {et~al.}(2008)\citenamefont {Shi},
  \citenamefont {Chang}, \citenamefont {Pailhes}, \citenamefont {Norman},
  \citenamefont {Campuzano}, \citenamefont {Mansson}, \citenamefont {Claesson},
  \citenamefont {Tjernberg}, \citenamefont {Bendounan}, \citenamefont
  {Patthey}, \citenamefont {Momono}, \citenamefont {Oda}, \citenamefont {Ido},
  \citenamefont {Mudry},\ and\ \citenamefont {Mesot}}]{Shi08}%
  \BibitemOpen
  \bibfield  {author} {\bibinfo {author} {\bibfnamefont {M.}~\bibnamefont
  {Shi}}, \bibinfo {author} {\bibfnamefont {J.}~\bibnamefont {Chang}}, \bibinfo
  {author} {\bibfnamefont {S.}~\bibnamefont {Pailhes}}, \bibinfo {author}
  {\bibfnamefont {M.~R.}\ \bibnamefont {Norman}}, \bibinfo {author}
  {\bibfnamefont {J.~C.}\ \bibnamefont {Campuzano}}, \bibinfo {author}
  {\bibfnamefont {M.}~\bibnamefont {Mansson}}, \bibinfo {author} {\bibfnamefont
  {T.}~\bibnamefont {Claesson}}, \bibinfo {author} {\bibfnamefont
  {O.}~\bibnamefont {Tjernberg}}, \bibinfo {author} {\bibfnamefont
  {A.}~\bibnamefont {Bendounan}}, \bibinfo {author} {\bibfnamefont
  {L.}~\bibnamefont {Patthey}}, \bibinfo {author} {\bibfnamefont
  {N.}~\bibnamefont {Momono}}, \bibinfo {author} {\bibfnamefont
  {M.}~\bibnamefont {Oda}}, \bibinfo {author} {\bibfnamefont {M.}~\bibnamefont
  {Ido}}, \bibinfo {author} {\bibfnamefont {C.}~\bibnamefont {Mudry}}, \ and\
  \bibinfo {author} {\bibfnamefont {J.}~\bibnamefont {Mesot}},\ }\href@noop {}
  {\bibfield  {journal} {\bibinfo  {journal} {Phys. Rev. Lett.}\ }\textbf
  {\bibinfo {volume} {101}},\ \bibinfo {pages} {047002} (\bibinfo {year}
  {2008})}\BibitemShut {NoStop}%
\bibitem [{\citenamefont {Shi}\ \emph {et~al.}(2009)\citenamefont {Shi},
  \citenamefont {Bendounan}, \citenamefont {Razzoli}, \citenamefont
  {Rosenkranz}, \citenamefont {Norman}, \citenamefont {Campuzano},
  \citenamefont {Chang}, \citenamefont {Mansson}, \citenamefont {Sassa},
  \citenamefont {Claesson}, \citenamefont {Tjernberg}, \citenamefont {Patthey},
  \citenamefont {Momono}, \citenamefont {Oda}, \citenamefont {Ido},
  \citenamefont {Guerrero}, \citenamefont {Mudry},\ and\ \citenamefont
  {Mesot}}]{Shi09}%
  \BibitemOpen
  \bibfield  {author} {\bibinfo {author} {\bibfnamefont {M.}~\bibnamefont
  {Shi}}, \bibinfo {author} {\bibfnamefont {A.}~\bibnamefont {Bendounan}},
  \bibinfo {author} {\bibfnamefont {E.}~\bibnamefont {Razzoli}}, \bibinfo
  {author} {\bibfnamefont {S.}~\bibnamefont {Rosenkranz}}, \bibinfo {author}
  {\bibfnamefont {M.~R.}\ \bibnamefont {Norman}}, \bibinfo {author}
  {\bibfnamefont {J.~C.}\ \bibnamefont {Campuzano}}, \bibinfo {author}
  {\bibfnamefont {J.}~\bibnamefont {Chang}}, \bibinfo {author} {\bibfnamefont
  {M.}~\bibnamefont {Mansson}}, \bibinfo {author} {\bibfnamefont
  {Y.}~\bibnamefont {Sassa}}, \bibinfo {author} {\bibfnamefont
  {T.}~\bibnamefont {Claesson}}, \bibinfo {author} {\bibfnamefont
  {O.}~\bibnamefont {Tjernberg}}, \bibinfo {author} {\bibfnamefont
  {L.}~\bibnamefont {Patthey}}, \bibinfo {author} {\bibfnamefont
  {N.}~\bibnamefont {Momono}}, \bibinfo {author} {\bibfnamefont
  {M.}~\bibnamefont {Oda}}, \bibinfo {author} {\bibfnamefont {M.}~\bibnamefont
  {Ido}}, \bibinfo {author} {\bibfnamefont {S.}~\bibnamefont {Guerrero}},
  \bibinfo {author} {\bibfnamefont {C.}~\bibnamefont {Mudry}}, \ and\ \bibinfo
  {author} {\bibfnamefont {J.}~\bibnamefont {Mesot}},\ }\href@noop {}
  {\bibfield  {journal} {\bibinfo  {journal} {Europhys. Lett.}\ }\textbf
  {\bibinfo {volume} {88}},\ \bibinfo {pages} {27008} (\bibinfo {year}
  {2009})}\BibitemShut {NoStop}%
\bibitem [{\citenamefont {Tanaka}\ \emph {et~al.}(2006)\citenamefont {Tanaka},
  \citenamefont {Lee}, \citenamefont {Lu}, \citenamefont {Fujimori},
  \citenamefont {Fujii}, \citenamefont {Risdiana}, \citenamefont {Terasaki},
  \citenamefont {Scalapino}, \citenamefont {Devereaux}, \citenamefont
  {Hussain},\ and\ \citenamefont {Shen}}]{Tanaka06}%
  \BibitemOpen
  \bibfield  {author} {\bibinfo {author} {\bibfnamefont {K.}~\bibnamefont
  {Tanaka}}, \bibinfo {author} {\bibfnamefont {W.~S.}\ \bibnamefont {Lee}},
  \bibinfo {author} {\bibfnamefont {D.~H.}\ \bibnamefont {Lu}}, \bibinfo
  {author} {\bibfnamefont {A.}~\bibnamefont {Fujimori}}, \bibinfo {author}
  {\bibfnamefont {T.}~\bibnamefont {Fujii}}, \bibinfo {author} {\bibnamefont
  {Risdiana}}, \bibinfo {author} {\bibfnamefont {I.}~\bibnamefont {Terasaki}},
  \bibinfo {author} {\bibfnamefont {D.~J.}\ \bibnamefont {Scalapino}}, \bibinfo
  {author} {\bibfnamefont {T.~P.}\ \bibnamefont {Devereaux}}, \bibinfo {author}
  {\bibfnamefont {Z.}~\bibnamefont {Hussain}}, \ and\ \bibinfo {author}
  {\bibfnamefont {Z.~X.}\ \bibnamefont {Shen}},\ }\href@noop {} {\bibfield
  {journal} {\bibinfo  {journal} {Science}\ }\textbf {\bibinfo {volume}
  {314}},\ \bibinfo {pages} {1910} (\bibinfo {year} {2006})}\BibitemShut
  {NoStop}%
\bibitem [{\citenamefont {Lee}\ \emph {et~al.}(2007)\citenamefont {Lee},
  \citenamefont {Vishik}, \citenamefont {Tanaka}, \citenamefont {Lu},
  \citenamefont {Sasagawa}, \citenamefont {Nagaosa}, \citenamefont {Devereaux},
  \citenamefont {Hussain},\ and\ \citenamefont {Shen}}]{Lee07}%
  \BibitemOpen
  \bibfield  {author} {\bibinfo {author} {\bibfnamefont {W.~S.}\ \bibnamefont
  {Lee}}, \bibinfo {author} {\bibfnamefont {I.~M.}\ \bibnamefont {Vishik}},
  \bibinfo {author} {\bibfnamefont {K.}~\bibnamefont {Tanaka}}, \bibinfo
  {author} {\bibfnamefont {D.~H.}\ \bibnamefont {Lu}}, \bibinfo {author}
  {\bibfnamefont {T.}~\bibnamefont {Sasagawa}}, \bibinfo {author}
  {\bibfnamefont {N.}~\bibnamefont {Nagaosa}}, \bibinfo {author} {\bibfnamefont
  {T.~P.}\ \bibnamefont {Devereaux}}, \bibinfo {author} {\bibfnamefont
  {Z.}~\bibnamefont {Hussain}}, \ and\ \bibinfo {author} {\bibfnamefont
  {Z.~X.}\ \bibnamefont {Shen}},\ }\href@noop {} {\bibfield  {journal}
  {\bibinfo  {journal} {Nature}\ }\textbf {\bibinfo {volume} {450}},\ \bibinfo
  {pages} {81} (\bibinfo {year} {2007})}\BibitemShut {NoStop}%
\bibitem [{\citenamefont {Terashima}\ \emph {et~al.}(2007)\citenamefont
  {Terashima}, \citenamefont {Matsui}, \citenamefont {Sato}, \citenamefont
  {Takahashi}, \citenamefont {Kofu},\ and\ \citenamefont
  {Hirota}}]{Terashima07}%
  \BibitemOpen
  \bibfield  {author} {\bibinfo {author} {\bibfnamefont {K.}~\bibnamefont
  {Terashima}}, \bibinfo {author} {\bibfnamefont {H.}~\bibnamefont {Matsui}},
  \bibinfo {author} {\bibfnamefont {T.}~\bibnamefont {Sato}}, \bibinfo {author}
  {\bibfnamefont {T.}~\bibnamefont {Takahashi}}, \bibinfo {author}
  {\bibfnamefont {M.}~\bibnamefont {Kofu}}, \ and\ \bibinfo {author}
  {\bibfnamefont {K.}~\bibnamefont {Hirota}},\ }\href@noop {} {\bibfield
  {journal} {\bibinfo  {journal} {Phys. Rev. Lett.}\ }\textbf {\bibinfo
  {volume} {99}},\ \bibinfo {pages} {017003} (\bibinfo {year}
  {2007})}\BibitemShut {NoStop}%
\bibitem [{\citenamefont {He}\ \emph {et~al.}(2009)\citenamefont {He},
  \citenamefont {Tanaka}, \citenamefont {Mo}, \citenamefont {Sasagawa},
  \citenamefont {Fujita}, \citenamefont {Adachi}, \citenamefont {Mannella},
  \citenamefont {Yamada}, \citenamefont {Koike}, \citenamefont {Hussain},\ and\
  \citenamefont {Shen}}]{He09}%
  \BibitemOpen
  \bibfield  {author} {\bibinfo {author} {\bibfnamefont {R.~H.}\ \bibnamefont
  {He}}, \bibinfo {author} {\bibfnamefont {K.}~\bibnamefont {Tanaka}}, \bibinfo
  {author} {\bibfnamefont {S.~K.}\ \bibnamefont {Mo}}, \bibinfo {author}
  {\bibfnamefont {T.}~\bibnamefont {Sasagawa}}, \bibinfo {author}
  {\bibfnamefont {M.}~\bibnamefont {Fujita}}, \bibinfo {author} {\bibfnamefont
  {T.}~\bibnamefont {Adachi}}, \bibinfo {author} {\bibfnamefont
  {N.}~\bibnamefont {Mannella}}, \bibinfo {author} {\bibfnamefont
  {K.}~\bibnamefont {Yamada}}, \bibinfo {author} {\bibfnamefont
  {Y.}~\bibnamefont {Koike}}, \bibinfo {author} {\bibfnamefont
  {Z.}~\bibnamefont {Hussain}}, \ and\ \bibinfo {author} {\bibfnamefont
  {Z.~X.}\ \bibnamefont {Shen}},\ }\href@noop {} {\bibfield  {journal}
  {\bibinfo  {journal} {Nature Phys.}\ }\textbf {\bibinfo {volume} {5}},\
  \bibinfo {pages} {119} (\bibinfo {year} {2009})}\BibitemShut {NoStop}%
\bibitem [{\citenamefont {Yoshida}\ \emph {et~al.}(2009)\citenamefont
  {Yoshida}, \citenamefont {Hashimoto}, \citenamefont {Ideta}, \citenamefont
  {Fujimori}, \citenamefont {Tanaka}, \citenamefont {Mannella}, \citenamefont
  {Hussain}, \citenamefont {Shen}, \citenamefont {Kubota}, \citenamefont {Ono},
  \citenamefont {Komiya}, \citenamefont {Ando}, \citenamefont {Eisaki},\ and\
  \citenamefont {Uchida}}]{Yoshida09}%
  \BibitemOpen
  \bibfield  {author} {\bibinfo {author} {\bibfnamefont {T.}~\bibnamefont
  {Yoshida}}, \bibinfo {author} {\bibfnamefont {M.}~\bibnamefont {Hashimoto}},
  \bibinfo {author} {\bibfnamefont {S.}~\bibnamefont {Ideta}}, \bibinfo
  {author} {\bibfnamefont {A.}~\bibnamefont {Fujimori}}, \bibinfo {author}
  {\bibfnamefont {K.}~\bibnamefont {Tanaka}}, \bibinfo {author} {\bibfnamefont
  {N.}~\bibnamefont {Mannella}}, \bibinfo {author} {\bibfnamefont
  {Z.}~\bibnamefont {Hussain}}, \bibinfo {author} {\bibfnamefont {Z.~X.}\
  \bibnamefont {Shen}}, \bibinfo {author} {\bibfnamefont {M.}~\bibnamefont
  {Kubota}}, \bibinfo {author} {\bibfnamefont {K.}~\bibnamefont {Ono}},
  \bibinfo {author} {\bibfnamefont {S.}~\bibnamefont {Komiya}}, \bibinfo
  {author} {\bibfnamefont {Y.}~\bibnamefont {Ando}}, \bibinfo {author}
  {\bibfnamefont {H.}~\bibnamefont {Eisaki}}, \ and\ \bibinfo {author}
  {\bibfnamefont {S.}~\bibnamefont {Uchida}},\ }\href@noop {} {\bibfield
  {journal} {\bibinfo  {journal} {Phys. Rev. Lett.}\ }\textbf {\bibinfo
  {volume} {103}},\ \bibinfo {pages} {037004} (\bibinfo {year}
  {2009})}\BibitemShut {NoStop}%
\bibitem [{\citenamefont {Kaminski}\ \emph {et~al.}(2004)\citenamefont
  {Kaminski}, \citenamefont {Rosenkranz}, \citenamefont {Fretwell},
  \citenamefont {Mesot}, \citenamefont {Randeria}, \citenamefont {Campuzano},
  \citenamefont {Norman}, \citenamefont {Li}, \citenamefont {Raffy},
  \citenamefont {Sato}, \citenamefont {Takahashi},\ and\ \citenamefont
  {Kadowaki}}]{Kaminski04}%
  \BibitemOpen
  \bibfield  {author} {\bibinfo {author} {\bibfnamefont {A.}~\bibnamefont
  {Kaminski}}, \bibinfo {author} {\bibfnamefont {S.}~\bibnamefont
  {Rosenkranz}}, \bibinfo {author} {\bibfnamefont {H.~M.}\ \bibnamefont
  {Fretwell}}, \bibinfo {author} {\bibfnamefont {J.}~\bibnamefont {Mesot}},
  \bibinfo {author} {\bibfnamefont {M.}~\bibnamefont {Randeria}}, \bibinfo
  {author} {\bibfnamefont {J.~C.}\ \bibnamefont {Campuzano}}, \bibinfo {author}
  {\bibfnamefont {M.~R.}\ \bibnamefont {Norman}}, \bibinfo {author}
  {\bibfnamefont {Z.~Z.}\ \bibnamefont {Li}}, \bibinfo {author} {\bibfnamefont
  {H.}~\bibnamefont {Raffy}}, \bibinfo {author} {\bibfnamefont
  {T.}~\bibnamefont {Sato}}, \bibinfo {author} {\bibfnamefont {T.}~\bibnamefont
  {Takahashi}}, \ and\ \bibinfo {author} {\bibfnamefont {K.}~\bibnamefont
  {Kadowaki}},\ }\href@noop {} {\bibfield  {journal} {\bibinfo  {journal}
  {Phys. Rev. B}\ }\textbf {\bibinfo {volume} {69}},\ \bibinfo {pages} {212509}
  (\bibinfo {year} {2004})}\BibitemShut {NoStop}%
\bibitem [{\citenamefont {Rosch}\ \emph {et~al.}(2005)\citenamefont {Rosch},
  \citenamefont {Gunnarsson}, \citenamefont {Zhou}, \citenamefont {Yoshida},
  \citenamefont {Sasagawa}, \citenamefont {Fujimori}, \citenamefont {Hussain},
  \citenamefont {Shen},\ and\ \citenamefont {Uchida}}]{Rosch05}%
  \BibitemOpen
  \bibfield  {author} {\bibinfo {author} {\bibfnamefont {O.}~\bibnamefont
  {Rosch}}, \bibinfo {author} {\bibfnamefont {O.}~\bibnamefont {Gunnarsson}},
  \bibinfo {author} {\bibfnamefont {X.~J.}\ \bibnamefont {Zhou}}, \bibinfo
  {author} {\bibfnamefont {T.}~\bibnamefont {Yoshida}}, \bibinfo {author}
  {\bibfnamefont {T.}~\bibnamefont {Sasagawa}}, \bibinfo {author}
  {\bibfnamefont {A.}~\bibnamefont {Fujimori}}, \bibinfo {author}
  {\bibfnamefont {Z.}~\bibnamefont {Hussain}}, \bibinfo {author} {\bibfnamefont
  {Z.~X.}\ \bibnamefont {Shen}}, \ and\ \bibinfo {author} {\bibfnamefont
  {S.}~\bibnamefont {Uchida}},\ }\href@noop {} {\bibfield  {journal} {\bibinfo
  {journal} {Phys. Rev. Lett.}\ }\textbf {\bibinfo {volume} {95}},\ \bibinfo
  {pages} {227002} (\bibinfo {year} {2005})}\BibitemShut {NoStop}%
\bibitem [{\citenamefont {Kim}\ \emph {et~al.}(2007)\citenamefont {Kim},
  \citenamefont {Park}, \citenamefont {Leem}, \citenamefont {Song},
  \citenamefont {Jin}, \citenamefont {Kim}, \citenamefont {Ronning},\ and\
  \citenamefont {Kim}}]{Kim07}%
  \BibitemOpen
  \bibfield  {author} {\bibinfo {author} {\bibfnamefont {C.}~\bibnamefont
  {Kim}}, \bibinfo {author} {\bibfnamefont {S.~R.}\ \bibnamefont {Park}},
  \bibinfo {author} {\bibfnamefont {C.~S.}\ \bibnamefont {Leem}}, \bibinfo
  {author} {\bibfnamefont {D.~J.}\ \bibnamefont {Song}}, \bibinfo {author}
  {\bibfnamefont {H.~U.}\ \bibnamefont {Jin}}, \bibinfo {author} {\bibfnamefont
  {H.~D.}\ \bibnamefont {Kim}}, \bibinfo {author} {\bibfnamefont
  {F.}~\bibnamefont {Ronning}}, \ and\ \bibinfo {author} {\bibfnamefont
  {C.}~\bibnamefont {Kim}},\ }\href@noop {} {\bibfield  {journal} {\bibinfo
  {journal} {Phys. Rev. B}\ }\textbf {\bibinfo {volume} {76}},\ \bibinfo
  {pages} {104505} (\bibinfo {year} {2007})}\BibitemShut {NoStop}%
\bibitem [{\citenamefont {Vishik}\ \emph {et~al.}(2010)\citenamefont {Vishik},
  \citenamefont {Lee}, \citenamefont {Schmitt}, \citenamefont {Moritz},
  \citenamefont {Sasagawa}, \citenamefont {Uchida}, \citenamefont {Fujita},
  \citenamefont {Ishida}, \citenamefont {Zhang}, \citenamefont {Devereaux},\
  and\ \citenamefont {Shen}}]{Vishik10}%
  \BibitemOpen
  \bibfield  {author} {\bibinfo {author} {\bibfnamefont {I.~M.}\ \bibnamefont
  {Vishik}}, \bibinfo {author} {\bibfnamefont {W.~S.}\ \bibnamefont {Lee}},
  \bibinfo {author} {\bibfnamefont {F.}~\bibnamefont {Schmitt}}, \bibinfo
  {author} {\bibfnamefont {B.}~\bibnamefont {Moritz}}, \bibinfo {author}
  {\bibfnamefont {T.}~\bibnamefont {Sasagawa}}, \bibinfo {author}
  {\bibfnamefont {S.}~\bibnamefont {Uchida}}, \bibinfo {author} {\bibfnamefont
  {K.}~\bibnamefont {Fujita}}, \bibinfo {author} {\bibfnamefont
  {S.}~\bibnamefont {Ishida}}, \bibinfo {author} {\bibfnamefont
  {C.}~\bibnamefont {Zhang}}, \bibinfo {author} {\bibfnamefont {T.~P.}\
  \bibnamefont {Devereaux}}, \ and\ \bibinfo {author} {\bibfnamefont {Z.~X.}\
  \bibnamefont {Shen}},\ }\href@noop {} {\bibfield  {journal} {\bibinfo
  {journal} {Phys. Rev. Lett.}\ }\textbf {\bibinfo {volume} {104}},\ \bibinfo
  {pages} {207002} (\bibinfo {year} {2010})}\BibitemShut {NoStop}%
\bibitem [{\citenamefont {Nakayama}\ \emph {et~al.}(2006)\citenamefont
  {Nakayama}, \citenamefont {Sato}, \citenamefont {Dobashi}, \citenamefont
  {Terashima}, \citenamefont {Souma}, \citenamefont {Matsui}, \citenamefont
  {Takahashi}, \citenamefont {Campuzano}, \citenamefont {Kudo}, \citenamefont
  {Sasaki}, \citenamefont {Kobayashi}, \citenamefont {Kondo}, \citenamefont
  {Takeuchi}, \citenamefont {Kadowaki}, \citenamefont {Kofu},\ and\
  \citenamefont {Hirota}}]{Nakayama06}%
  \BibitemOpen
  \bibfield  {author} {\bibinfo {author} {\bibfnamefont {K.}~\bibnamefont
  {Nakayama}}, \bibinfo {author} {\bibfnamefont {T.}~\bibnamefont {Sato}},
  \bibinfo {author} {\bibfnamefont {T.}~\bibnamefont {Dobashi}}, \bibinfo
  {author} {\bibfnamefont {K.}~\bibnamefont {Terashima}}, \bibinfo {author}
  {\bibfnamefont {S.}~\bibnamefont {Souma}}, \bibinfo {author} {\bibfnamefont
  {H.}~\bibnamefont {Matsui}}, \bibinfo {author} {\bibfnamefont
  {T.}~\bibnamefont {Takahashi}}, \bibinfo {author} {\bibfnamefont {J.~C.}\
  \bibnamefont {Campuzano}}, \bibinfo {author} {\bibfnamefont {K.}~\bibnamefont
  {Kudo}}, \bibinfo {author} {\bibfnamefont {T.}~\bibnamefont {Sasaki}},
  \bibinfo {author} {\bibfnamefont {N.}~\bibnamefont {Kobayashi}}, \bibinfo
  {author} {\bibfnamefont {T.}~\bibnamefont {Kondo}}, \bibinfo {author}
  {\bibfnamefont {T.}~\bibnamefont {Takeuchi}}, \bibinfo {author}
  {\bibfnamefont {K.}~\bibnamefont {Kadowaki}}, \bibinfo {author}
  {\bibfnamefont {M.}~\bibnamefont {Kofu}}, \ and\ \bibinfo {author}
  {\bibfnamefont {K.}~\bibnamefont {Hirota}},\ }\href@noop {} {\bibfield
  {journal} {\bibinfo  {journal} {Phys. Rev. B}\ }\textbf {\bibinfo {volume}
  {74}},\ \bibinfo {pages} {054505} (\bibinfo {year} {2006})}\BibitemShut
  {NoStop}%
\bibitem [{\citenamefont {King}\ \emph {et~al.}(2011)\citenamefont {King},
  \citenamefont {Rosen}, \citenamefont {Meevasana}, \citenamefont {Tamai},
  \citenamefont {Rozbicki}, \citenamefont {Comin}, \citenamefont {Levy},
  \citenamefont {Fournier}, \citenamefont {Yoshida}, \citenamefont {Eisaki},
  \citenamefont {Shen}, \citenamefont {Ingle}, \citenamefont {Damascelli},\
  and\ \citenamefont {Baumberger}}]{King11}%
  \BibitemOpen
  \bibfield  {author} {\bibinfo {author} {\bibfnamefont {P.~D.~C.}\
  \bibnamefont {King}}, \bibinfo {author} {\bibfnamefont {J.~A.}\ \bibnamefont
  {Rosen}}, \bibinfo {author} {\bibfnamefont {W.}~\bibnamefont {Meevasana}},
  \bibinfo {author} {\bibfnamefont {A.}~\bibnamefont {Tamai}}, \bibinfo
  {author} {\bibfnamefont {E.}~\bibnamefont {Rozbicki}}, \bibinfo {author}
  {\bibfnamefont {R.}~\bibnamefont {Comin}}, \bibinfo {author} {\bibfnamefont
  {G.}~\bibnamefont {Levy}}, \bibinfo {author} {\bibfnamefont {D.}~\bibnamefont
  {Fournier}}, \bibinfo {author} {\bibfnamefont {Y.}~\bibnamefont {Yoshida}},
  \bibinfo {author} {\bibfnamefont {H.}~\bibnamefont {Eisaki}}, \bibinfo
  {author} {\bibfnamefont {K.~M.}\ \bibnamefont {Shen}}, \bibinfo {author}
  {\bibfnamefont {N.~J.~C.}\ \bibnamefont {Ingle}}, \bibinfo {author}
  {\bibfnamefont {A.}~\bibnamefont {Damascelli}}, \ and\ \bibinfo {author}
  {\bibfnamefont {F.}~\bibnamefont {Baumberger}},\ }\href@noop {} {\bibfield
  {journal} {\bibinfo  {journal} {Phys. Rev. Lett.}\ }\textbf {\bibinfo
  {volume} {106}},\ \bibinfo {pages} {127005} (\bibinfo {year}
  {2011})}\BibitemShut {NoStop}%
\bibitem [{\citenamefont {Norman}\ \emph {et~al.}(1998)\citenamefont {Norman},
  \citenamefont {Randeria}, \citenamefont {Ding},\ and\ \citenamefont
  {Campuzano}}]{Norman98}%
  \BibitemOpen
  \bibfield  {author} {\bibinfo {author} {\bibfnamefont {M.~R.}\ \bibnamefont
  {Norman}}, \bibinfo {author} {\bibfnamefont {M.}~\bibnamefont {Randeria}},
  \bibinfo {author} {\bibfnamefont {H.}~\bibnamefont {Ding}}, \ and\ \bibinfo
  {author} {\bibfnamefont {J.~C.}\ \bibnamefont {Campuzano}},\ }\href@noop {}
  {\bibfield  {journal} {\bibinfo  {journal} {Phys. Rev. B}\ }\textbf {\bibinfo
  {volume} {57}},\ \bibinfo {pages} {11093 (R)} (\bibinfo {year}
  {1998})}\BibitemShut {NoStop}%
\bibitem [{\citenamefont {McElroy}\ \emph {et~al.}(2005)\citenamefont
  {McElroy}, \citenamefont {Lee}, \citenamefont {Hoffman}, \citenamefont
  {Lang}, \citenamefont {Lee}, \citenamefont {Hudson}, \citenamefont {Eisaki},
  \citenamefont {Uchida},\ and\ \citenamefont {Davis}}]{McElroy05}%
  \BibitemOpen
  \bibfield  {author} {\bibinfo {author} {\bibfnamefont {K.}~\bibnamefont
  {McElroy}}, \bibinfo {author} {\bibfnamefont {D.~H.}\ \bibnamefont {Lee}},
  \bibinfo {author} {\bibfnamefont {J.~E.}\ \bibnamefont {Hoffman}}, \bibinfo
  {author} {\bibfnamefont {K.~M.}\ \bibnamefont {Lang}}, \bibinfo {author}
  {\bibfnamefont {J.}~\bibnamefont {Lee}}, \bibinfo {author} {\bibfnamefont
  {E.~W.}\ \bibnamefont {Hudson}}, \bibinfo {author} {\bibfnamefont
  {H.}~\bibnamefont {Eisaki}}, \bibinfo {author} {\bibfnamefont
  {S.}~\bibnamefont {Uchida}}, \ and\ \bibinfo {author} {\bibfnamefont {J.~C.}\
  \bibnamefont {Davis}},\ }\href@noop {} {\bibfield  {journal} {\bibinfo
  {journal} {Physical Review Letters}\ }\textbf {\bibinfo {volume} {94}},\
  \bibinfo {pages} {197005} (\bibinfo {year} {2005})}\BibitemShut {NoStop}%
\bibitem [{\citenamefont {Pushp}\ \emph {et~al.}(2009)\citenamefont {Pushp},
  \citenamefont {Parker}, \citenamefont {Pasupathy}, \citenamefont {Gomes},
  \citenamefont {Ono}, \citenamefont {Wen}, \citenamefont {Xu}, \citenamefont
  {Gu},\ and\ \citenamefont {Yazdani}}]{Pushp09}%
  \BibitemOpen
  \bibfield  {author} {\bibinfo {author} {\bibfnamefont {A.}~\bibnamefont
  {Pushp}}, \bibinfo {author} {\bibfnamefont {C.~V.}\ \bibnamefont {Parker}},
  \bibinfo {author} {\bibfnamefont {A.~N.}\ \bibnamefont {Pasupathy}}, \bibinfo
  {author} {\bibfnamefont {K.~K.}\ \bibnamefont {Gomes}}, \bibinfo {author}
  {\bibfnamefont {S.}~\bibnamefont {Ono}}, \bibinfo {author} {\bibfnamefont
  {J.~S.}\ \bibnamefont {Wen}}, \bibinfo {author} {\bibfnamefont {Z.~J.}\
  \bibnamefont {Xu}}, \bibinfo {author} {\bibfnamefont {G.}~\bibnamefont {Gu}},
  \ and\ \bibinfo {author} {\bibfnamefont {A.}~\bibnamefont {Yazdani}},\
  }\href@noop {} {\bibfield  {journal} {\bibinfo  {journal} {Science}\ }\textbf
  {\bibinfo {volume} {324}},\ \bibinfo {pages} {1689} (\bibinfo {year}
  {2009})}\BibitemShut {NoStop}%
\end{thebibliography}

\end{document}